\documentclass[pra,aps,amsmath,amssymb,twocolumn,groupedaddress ,floatfix,hidelinks]{revtex4-2}
\usepackage{graphicx,
            amsmath,
            braket,
            units,
            ragged2e,
            xcolor,
            xspace,
            nicefrac,
            lipsum,
            nameref,
            textcomp,
            urwchancal,
            upgreek,
            isotope,
            hyperref,
            cleveref,
            placeins, 
            scrextend, 
            url}
\usepackage{amstext}
\usepackage{amsfonts}
\usepackage{amsxtra}
\usepackage{bm}
\usepackage{grffile}
\usepackage{soul}
\usepackage{epstopdf}
\usepackage{marvosym} 
\usepackage{tikz}
\usepackage{flushend}
\usepackage{scalerel}

\definecolor{ultramarine}{rgb}{0.07, 0.04, 0.56}

\usetikzlibrary{svg.path}

\hypersetup{
    colorlinks=true,
    allcolors = blue
}

\DeclareMathAlphabet{\mathcal}{OMS}{cmsy}{m}{n}

{%
\setlength{\fboxsep}{0pt}%
\setlength{\fboxrule}{1pt}%
}%

\definecolor{orcidlogocol}{HTML}{A6CE39}
\tikzset{
  orcidlogo/.pic={
    \fill[orcidlogocol] svg{M256,128c0,70.7-57.3,128-128,128C57.3,256,0,198.7,0,128C0,57.3,57.3,0,128,0C198.7,0,256,57.3,256,128z};
    \fill[white] svg{M86.3,186.2H70.9V79.1h15.4v48.4V186.2z}
                 svg{M108.9,79.1h41.6c39.6,0,57,28.3,57,53.6c0,27.5-21.5,53.6-56.8,53.6h-41.8V79.1z M124.3,172.4h24.5c34.9,0,42.9-26.5,42.9-39.7c0-21.5-13.7-39.7-43.7-39.7h-23.7V172.4z}
                 svg{M88.7,56.8c0,5.5-4.5,10.1-10.1,10.1c-5.6,0-10.1-4.6-10.1-10.1c0-5.6,4.5-10.1,10.1-10.1C84.2,46.7,88.7,51.3,88.7,56.8z};
  }
}

\newcommand\orcidicon[1]{\href{https://orcid.org/#1}{\mbox{\scalerel*{
\begin{tikzpicture}[yscale=-1,transform shape]
\pic{orcidlogo};
\end{tikzpicture}
}{|}}}}

\begin{document}

\title{Anomalous dispersion of shear waves in dipolar supersolids}

\author{P. Senarath Yapa\,\orcidicon{0000-0002-5031-4695}}
\email{pramodh.sy@gmail.com}
\affiliation{
Universität Innsbruck, Fakultät für Mathematik, Informatik und Physik,
Institut für Experimentalphysik, 6020 Innsbruck, Austria
}

\author{T. Bland\,\orcidicon{0000-0001-9852-0183}}
\email{thomas.bland1991@gmail.com}
\affiliation{
Universität Innsbruck, Fakultät für Mathematik, Informatik und Physik,
Institut für Experimentalphysik, 6020 Innsbruck, Austria
}

\begin{abstract}
Dipolar supersolids---quantum states that are simultaneously superfluid and solid---have had their superfluid nature rigorously tested, while its solid nature remains uncharted. Arguably, the defining characteristic of a solid is the existence of elastic shear waves. In this work, we investigate transverse wave packet propagation in dipolar supersolids with triangular and honeycomb structure. Remarkably, the honeycomb supersolid displays anomalous dispersion, supporting waves traveling faster than the transverse speed of sound. While the distinction between anomalous and normal dispersion appears subtle, it has a profound influence on both the transport and scattering properties. For both supersolid phases, we calculate the shear modulus, a key parameter that quantifies the material's rigidity. Our findings are pertinent to current experimental efforts scrutinizing the fundamental properties of supersolids.
\end{abstract}

\date{\today}
\maketitle

Shear waves---an elastic wave typified by oscillation of the medium perpendicular to the direction of wave propagation---are an invaluable tool for probing the internal structures of various systems over a range of scales. In geophysics, they are used to reveal the Earth's layered structure \cite{Lehmann1936p}; in medical imaging, shear wave elastography techniques characterize tissue composition \cite{Sarvazyan1998swe}; and down at the quantum level, they are used to probe the elasticity of Abrikosov vortex lattices in superfluids \cite{Tkachenko1966ovl,Tkachenko1966sov,Tkachenko1969eov,Coddington2003oot}.

One other such quantum phase that should support shear waves is the supersolid. A supersolid is a fascinating and counterintuitive state of matter that combines the properties of superfluidity with crystalline order \cite{Gross1957uto,Gross1958cto,Andreev1969qto,Thouless1969tfo,Chester1970sob,Leggett1970cas}. In this exotic phase, a material exhibits both the rigidity and periodic structure characteristic of an ordered solid, and the inviscid flow typically associated with superfluids. This duality emerges from multiple spontaneously broken symmetries: a global U(1) phase symmetry, and both translation and rotation symmetries \cite{NIELSEN1976_goldstonemodes}. Ultracold atoms have now established themselves as the leading platform to study supersolidity, with varied approaches including cavity settings \cite{Leonard2017sfi,leonard2017mam,guo2021aol}, atoms with spin-orbit coupling \cite{Li2017asp}, and dipolar atoms exhibiting intrinsic long-range interactions \cite{Boettcher2019tsp,Tanzi2019ooa,Chomaz2019lla,Norcia2021tds}. Understanding this phase in the context of ultracold matter may unveil new behavior in solid state materials \cite{Xiang2024gme}, helium systems \cite{sridhar1987apd,Nyeki2017isa,Levitin2019efa,Shook2020spd,Yapa2022}, and neutron stars \cite{Chamel2012nci,Poli2023gir,Bland2024epg}.

Experiments on supersolids composed of dipolar atoms have opened numerous avenues to probe their superfluid nature \cite{Chomaz2022dpa,Recati2023siu}, including measurements of the global phase coherence \cite{Boettcher2019tsp,Tanzi2019ooa,Chomaz2019lla}, the existence of two phononic branches in one-dimensional crystals (one each for the broken phase and translation symmetries) \cite{Guo2019tle,Natale2019eso,Tanzi2019ssb}, Josephson-type dynamics \cite{Ilzhoefer2021pci,Biagioni2024mot}, and most recently quantum vortices \cite{Casotti2024oov}. Though the solid nature is inferred from the crystalline density distribution and the corresponding Bragg peaks at finite momenta \cite{Boettcher2019tsp,Tanzi2019ooa,Chomaz2019lla,Petter2021bso}, fundamental elastic properties associated with solids, such as rigidity, stress, strain, and shear wave propagation, have yet to be investigated. This leads to a natural question: How solid is a dipolar supersolid? \cite{Kuklov2011hsi}.


Here, we propose a protocol to excite shear waves in two-dimensional dipolar supersolids, and measure their resulting propagation velocities. We investigate this protocol for triangular and honeycomb lattice structures, which are ground state solutions for a system of ultracold dipolar atoms under planar confinement \cite{Zhang2019saa,Ripley2023tds,Zhang2024mpi}. Our results reveal distinct dispersive phenomena between the triangular and honeycomb phases, which we probe through the current density dynamic structure factor (DSF) and the Bogoliubov de-Gennes (BdG) method. These investigations show that the honeycomb structure can support a refractive index smaller than unity, thus supporting shear waves traveling faster than the transverse speed of sound. The idea of such a sound mode in classical wave theory can be traced back to Sommerfeld and  Brillouin~\cite{sommerfeld1907,Sommerfeld1914,Brillouin1914,Brillouin:1960tos}, while in the quantum realm, it is analogous to the anomalous phonon dispersion relation observed in superfluid helium-4 \cite{cowley1971iso,dynes1975moa,Maris1977ppi,sridhar1987apd}. We speculate that this anomalous behavior is a signature of an incipient transverse branch instability in the honeycomb supersolid as recently reported by Blakie~\cite{Blakie2025}. Finally, we find the relationship between the wave packet propagation velocities and the transverse speed of sound, and estimate the shear modulus of dipolar supersolids.

\begin{figure*}[!t]
    \centering
    \includegraphics[width=1\linewidth]{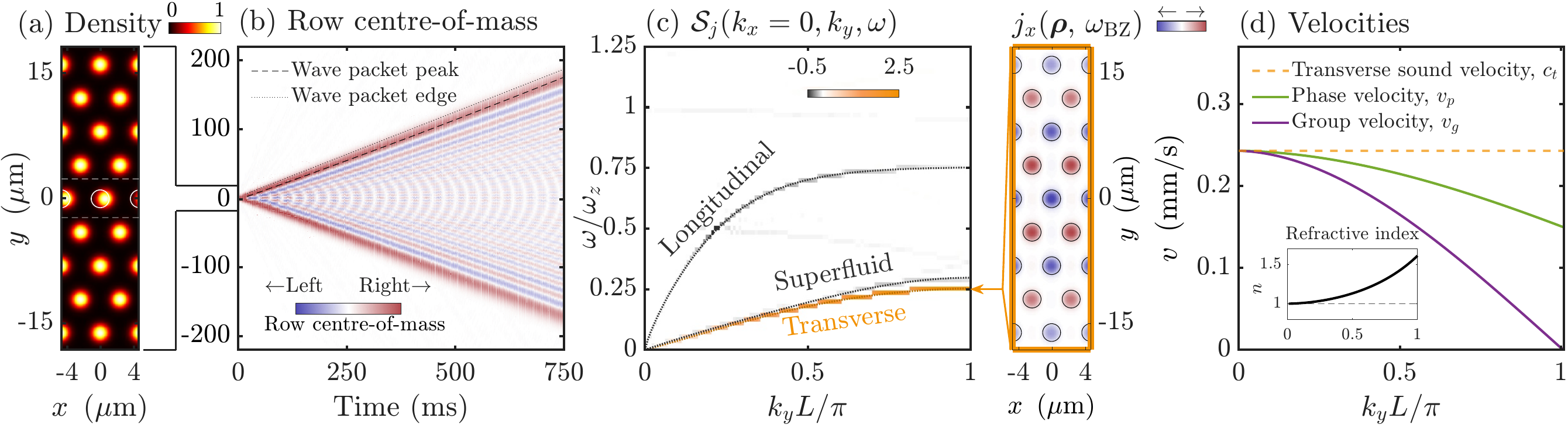}
    \caption{Excitation of shear waves in a triangular lattice dipolar supersolid, with $(a_s,n_\text{2D}) = (0.75\,a_\text{dd}, 1500\,\mu$m$^{-2})$. (a) The initial density normalized to the peak density, which has been locally perturbed along $\hat{x}$ (between the dashed lines). The white circles show the initial droplet positions. (b) Shear wave propagation along $\hat y$ shown by the deviation of the row COM over time. (c) Current density DSF (Eq.~\eqref{eqn:cDSF}) and the gapless branches obtained via BdG calculations (dotted lines). The dominant fluctuation in $j_x$ at $\omega_\text{BZ}=0.251 \omega_z$ is highlighted, with contours shown at 30\% of the maximum density. (d) Transverse velocities, calculated from the transverse branch of the BdG results shown in (c). Inset shows the refractive index $n(k_y)$.}
    \label{fig:tri}
\end{figure*}

{\it Theoretical framework}---In our investigations, we study a dilute weakly-interacting gas of ultracold dipolar atoms in a Bose Einstein condensate (BEC). Under a homogeneous magnetic field, the interactions can be modelled by a simple two-body pseudo-potential
\begin{eqnarray}
    U(\textbf{r}) = \frac{4\pi\hbar^2a_{\mathrm s}}{m}\delta(\textbf{r})+\frac{3\hbar^2a_\text{dd}}{m}\frac{1-3\cos^2\theta}{r^3}\,,
    \label{eqn:U}
\end{eqnarray}
for particles with mass $m$. The nature of the short-range contact interactions are governed by the size and sign of the s-wave scattering length $a_\text{s}$---tuneable through Feshbach resonances \cite{Chin2010fri}---and the long-range anisotropic dipole-dipole interactions are similarly governed by the fixed dipolar scattering length $a_\text{dd}$. The dipoles are polarized along the $z$ axis and $\theta$ is the angle between $\textbf{r}$ and the $z$ axis. Here, we focus our studies on $^{164}$Dy atoms in the bosonic ground state, where $a_\text{dd}=130.8\,a_0$ with Bohr radius $a_0$, though they are generally applicable to other atomic species.

This full system is well described by a beyond mean-field theory, known as the extended Gross-Pitaevskii equation (eGPE) \cite{Waechtler2016qfi,FerrierBarbut2016ooq, Chomaz2016qfd,Bisset2016gsp}. This governs the three-dimensional macroscopic wavefunction $\Psi(\textbf{r},t)$, and takes into account single-particle dynamics, two-body interactions, and quantum fluctuations. Due to the azimuthal symmetry of the interaction, dipolar supersolids break at most two translational symmetries, giving rise to triangular, stripe, or honeycomb two-dimensional (2D) lattice structures \cite{Zhang2019saa,Ripley2023tds,Zhang2024mpi}. Thus, following Ref.~\cite{Ripley2023tds}, we are able to reduce the complexity of this model to the quasi-2D variational eGPE
\begin{align}
    i\hbar\frac{\partial\psi}{\partial t} =  \bigg[-\frac{\hbar^2\nabla^2}{2m} + \Phi_\sigma+g_\text{QF}|\psi|^3 \bigg]\psi\,,
    \label{eqn:2DGPE}
\end{align}
where the dimensionally reduced wave function $\psi(\boldsymbol{\rho},t)$ is defined over the infinite plane, $\boldsymbol{\rho}=(x,y)$. The 2D interaction and quantum fluctuation terms, $\Phi_\sigma$ and $g_\text{QF}$, depend on the variational width $\sigma$ in the perpendicular ($z$) direction. We fix the areal number density $n_\text{2D}$, and perpendicular harmonic trap with frequency $\omega_z =2 \pi\times72.4 \mathrm{\,Hz}$. This model and our numerical methods are detailed in the Supplemental Material~\cite{supp}. 

\begin{figure*}[!t]
    \centering
    \includegraphics[width=1\linewidth]{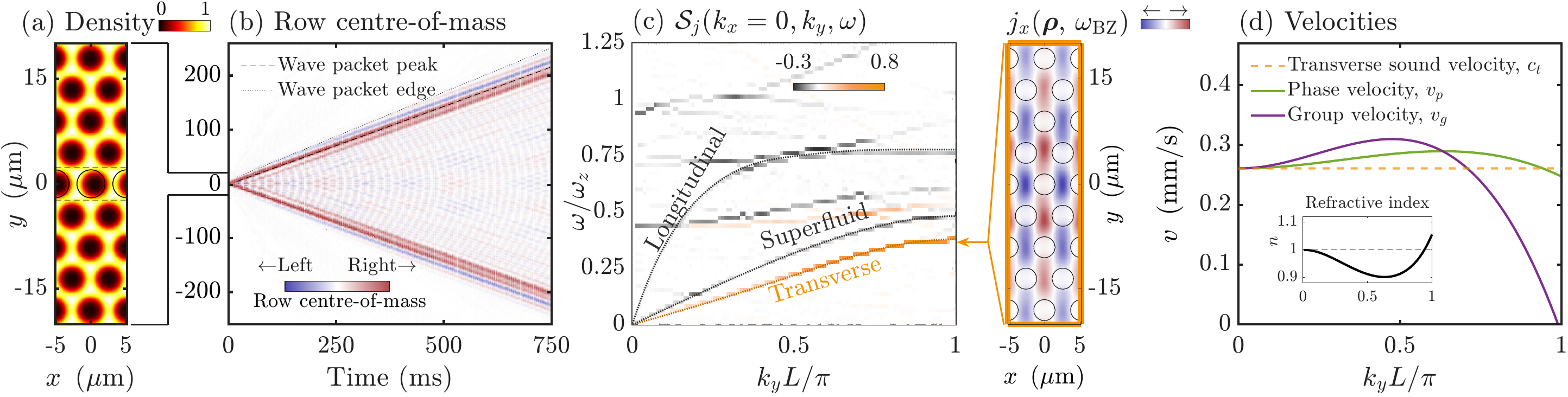}
    \caption{Excitation of shear waves in a honeycomb lattice dipolar supersolid, with $(a_s,n_\text{2D}) = (0.775\,a_\text{dd}, 4000\,\mu$m$^{-2})$. Panels (a)-(d) as in Fig.~\ref{fig:tri}, but with $\omega_\text{BZ}=0.366\omega_z$.}
    \label{fig:hex}
\end{figure*}

The ground state phases of this variational quasi-2D model can host supersolids with intriguing density structures. When $a_s>a_\text{dd}$, the state can only be a uniform planar gas. The dipolar interactions are not strong enough to modulate the density, however, there can be a roton minimum in the excitation spectrum \cite{ODell2003rig,Santos2003rms,Chomaz2018oor,Schmidt2021rei}. In the dipolar-dominated regime, $a_s<a_\text{dd}$, the roton gap can close and the system spontaneously breaks the spatial symmetries \cite{Recati2023siu}. Of the available density modulated states, two ground state configurations can carry shear waves. At low densities, the system crystallizes into a triangular pattern, as recently observed in finite sized systems \cite{Norcia2021tds,Bland2022tds}. At large densities, the system inverts, giving rise to a triangular arrangement of holes and a honeycomb lattice structure \cite{Zhang2019saa}. There is another ground state configuration at intermediate densities with one broken translational symmetry---the stripe phase \cite{Ripley2023tds}---that does not support transverse excitations \footnote{There are also non-ground state (metastable) two-dimensional structures predicted \cite{Zhang2024mpi} that we do not consider here.}.

{\it Shear waves in triangular supersolids}---In Fig.~\ref{fig:tri}(a), we show the initial perturbation of a single row of the crystal away from its equilibrium position, which adds elastic tension to the lattice. We shift the central row by $\Delta L = 0.125 L$, where $L$ is the lattice spacing. We have ensured that the observed shear wave propagation does not vary with the magnitude of this shift for smaller perturbations. Evolving the system in real time, we observe a longitudinally propagating wave packet (along $\hat{y}$), leaving behind transverse crystal oscillations (along $\hat{x}$). We then track the deviation of the center of mass (COM) of each row from its equilibrium position, as shown in Fig.~\ref{fig:tri}(b). We track both the peak and edge of this wave packet, defining the latter as a COM deviation larger than 5\% of the peak amplitude. The peak velocity increases with the propagation time, starting at $\approx0.23\,$mm/s and grows to $\approx0.24\,$mm/s after 750 ms, and the edge velocity always exceeds the peak velocity at $\approx0.26\,$mm/s, due to the dispersive nature of the wave packet. The existence of shear waves implies a non-zero shear modulus of dipolar supersolids, and proves the solid-like nature of the state, a key result of this work.

We can begin to discern the accelerating wave packet dynamics from the dispersion relation, obtained by calculating the current density DSF. Given the current density $\boldsymbol{j} = (j_x,j_y)$, calculated through
${
\boldsymbol{j}(\boldsymbol{\rho},t)=\frac{\hbar}{2 im}\left[ \psi^*(\boldsymbol{\rho},t) \nabla  \psi(\boldsymbol{\rho},t)- \psi(\boldsymbol{\rho},t) \nabla  \psi^*(\boldsymbol{\rho},t)\right]\,,
}$
we can define the current density DSF as ${\mathcal{S}_j = \mathcal{S}_{j_x} - \mathcal{S}_{j_y}}$, where 
\begin{equation}
\mathcal{S}_{j_\alpha}(\boldsymbol{k},\omega)=\left|\iint \mathrm{d} \boldsymbol{\rho}\,\mathrm{d} t\,e^{-i \omega t+i \boldsymbol{k}\cdot\boldsymbol{\rho}}j_\alpha\right|\,,
\label{eqn:cDSF}
\end{equation}
reminiscent of the transverse DSF defined in Ref.~\cite{Tomoshige2021}. In this definition, $\mathcal{S}_j>0$ corresponds to a transverse current density excitation ($j_x$) and $\mathcal{S}_j<0$ corresponds to a longitudinal current density excitation ($j_y$). The amplitude of $\mathcal{S}_j$ gives the strength of the response of the corresponding $(\boldsymbol{k},\omega)$ fluctuation to the excitation.  

Fig.~\ref{fig:tri}(c) shows half of the first Brillouin Zone (BZ), $0<k_y<k_\text{BZ}=\pi/L$, of the current density DSF, revealing the three gapless (Goldstone) modes inherent to a 2D supersolid \cite{Watanabe2012sbo}. As apparent from the large amplitude positive $\mathcal{S}_j$, the response of the system is dominated by $j_x$ fluctuations, revealing the transverse branch $\omega_t(k_y)$. Only weak longitudinal crystal and superfluid phonons are induced, as expected from the dislocation in the density at the boundary of the initial perturbation. On the right-hand side of (c), we also highlight the dominant current density fluctuation by taking $\mathcal{F}_t[j_x(\boldsymbol{\rho},t)] = j_x(\boldsymbol{\rho},\omega)$, and isolating the mode with frequency $\omega_t(k_\text{BZ})=\omega_\text{BZ}$ \cite{Norcia2022cao}. We note that $\omega_\text{BZ}$ is equal to the oscillation frequency of the central row COM ($y=0$), shown in (b). This mode corresponds to a transverse phonon with wave number $k_\text{BZ}$, i.e.~a fluctuation where the current alternates between each row.

We independently verify the dispersion of these Goldstone modes by introducing a perturbation about the ground state solution, $\psi(\boldsymbol{\rho},t)=\left[\psi_0(\boldsymbol{\rho})+\delta (\boldsymbol{\rho},t)\right]\mathrm{e}^{-\mathrm{i} \mu t / \hbar}$. Here, $\mu$ is the chemical potential and 

\begin{equation}\label{eqn:pert}
\delta (\boldsymbol{\rho},t)=\sum_{\nu}\left[u_\nu(\boldsymbol{\rho}) \mathrm{e}^{-\mathrm{i} \omega_\nu t}-v_\nu^*(\boldsymbol{\rho}) \mathrm{e}^{\mathrm{i} \omega_\nu^* t}\right],
\end{equation}
where $u_\nu$ and $v_\nu$ are the Bogoliubov modes with band index $\nu$ which obey the translational symmetry of the supersolid state. By substituting the perturbed wavefunction into Eq.~\eqref{eqn:2DGPE} and linearizing, we obtain the familiar Bogoliubov-de Gennes (BdG) equations~\cite{Macri2013,Poli2024eoa} (see the Supplemental Material for details~\cite{supp}). We solve the BdG equations using a Bloch wave expansion of the Bogoliubov modes to obtain the frequencies, $\omega_\nu(\boldsymbol{k})$, and overlay the three lowest bands for comparison in Fig.~\ref{fig:tri}(c). Note that the near-identical shape of the superfluid and transverse modes is a coincidence of the chosen parameters, and their speeds of sound are distinct in other regions of parameter space~\cite{Poli2024eoa}. We can extract three characteristic velocities from the BdG results for the transverse branch: 1) the transverse speed of sound $c_t=\lim_{k_y\rightarrow0}\omega_t/k_y$, 2) the phase velocity $v_p(k_y) =  \omega_t/ k_y$, which defines the velocity of a wave with given $k_y$, and 3) the group velocity, $v_g(k_y) = \partial \omega_t/\partial k_y$, which defines the velocity of the wave packet envelope and thus the velocity of energy propagation in the medium \footnote{We extract these quantities from the BdG calculations by fitting the transverse branch with a fourth-order polynomial.}. For the triangular supersolid, both $v_g$ and $v_p$ monotonically decrease from $c_t$ with increasing wave number, and subsequently the refractive index, $n = c_t/v_p$, increases from unity with $k_y$. This behavior is referred to as \textit{normal dispersion} and is a hallmark of conventional crystalline solids \cite{kittel2018introduction}.

The $k$-dependence of these characteristic velocities fully describes the observed shear wave packet dynamics in Fig.~\ref{fig:tri}(b). The initial perturbation is localized in space, exciting a range of high-$k$ modes along the transverse branch. The group velocity of this wave packet is initially smaller than the speed of sound, increasing in velocity as the wave packet broadens and involves the lower-$k$ modes. The wave packet edge travels at the largest $v_g$, which in this case corresponds to $c_t$. Consistent with this interpretation, we have verified that initially perturbing multiple crystal rows gives qualitatively similar results, but with a larger initial peak velocity due to a greater coupling with long wavelength low-$k$ modes. As the wave packet moves across the system, only the mode with $v_g=0$ is left behind, corresponding to the mode at the BZ edge, highlighted in Fig.~\ref{fig:tri}(c).

{\it Anomalous dispersion of shear waves in honeycomb supersolids}---Applying the same protocol to the honeycomb supersolid [Fig.~\ref{fig:hex}(a)], we observe radically distinct behavior. For the triangular supersolid, the wave packet edge propagates at the speed of sound, where $\omega=0$, as indicated by an enduring rightward COM displacement (colored red in Fig.~\ref{fig:tri}(b)). In contrast, the wave packet edge in the honeycomb supersolid oscillates between left and right displacement [Fig.~\ref{fig:hex}(b)]. Thus, the fastest velocity in the excitation has $\omega>0$, and exceeds the speed of sound.

We can pinpoint the origin of this unexpected finding from the current density DSF, the BdG dispersion and the corresponding characteristic velocities, as shown in Figs.~\ref{fig:hex}(c)-(d) \footnote{Note that as the perturbation dislocates the high density region, we excite many branches, but still dominantly transverse crystal modes.}. The transverse branch displays a subtle upward concavity for intermediate wave vector ($0.2 \lesssim k_yL/\pi\lesssim
0.4$), supporting $v_g$ and $v_p$ which exceed $c_t$, and a sub-unity refractive index. This is known as {\it anomalous dispersion} and has significant implications for transverse phonon decay. The superlinear dispersion allows for Beliaev damping of transverse phonons, where one phonon spontaneously decays into two \cite{Beliaev1958aot,Beliaev1958eso}. Though rare, similar phenomena has been observed in other systems such as crystalline materials \cite{Baumgartner1981sdo,Maris1985sdo}, Wigner and Coulomb crystals~\cite{wang2001lat,peeters1987wco,platzman1974pdo}  and superfluid helium-4 \cite{dynes1975moa}, the latter being the subject of thirty years of intense experimental and theoretical debate \cite{sridhar1987apd}.  

The origin of the anomalous behavior can be understood through a simple physical argument, though this will be substantiated in a subsequent work. In general, the dispersion relation for a gapless branch must have the form $\omega(k) = Ak + Bk^2+\mathcal{O}(k^3)$ for momentum $k$. A phonon instability occurs when the linear part becomes negative, crossing $A=0$. However, in a crystalline system at the Brillouin zone (BZ) edge $\text{d}\omega(k)/\text{d}k = 0$. Therefore, near to the shear wave (transverse phonon) instability the band goes as $k^2$ for low momenta and flattens at the BZ edge. These criteria set ``boundary conditions" for the band and lead to anomalous behavior for intermediate $k$. We note that this implies an enhancement of the anomalous behavior close to the shear wave instability~\cite{Blakie2025}, however, the honeycomb state is not the ground state in that region, and it is not be possible to dynamically probe this in an experiment without entering the stripe regime. This was shown in the top row of Fig.~3 in the very recent work of Ref.~\cite{Blakie2025}. Our results show that this behavior persists in an experimentally relevant parameter regime where the honeycomb state is indeed the ground state. To our knowledge, the honeycomb supersolid constitutes the first quantum solid to display anomalous dispersion, the main result of this Letter.

{\it Estimating the shear modulus}---There is a simple expression that relates the shear wave velocity $v_s$ and the material density to the shear modulus $\mu$---the measure of elastic shear stiffness of a medium---given by \cite{Andreev1969qto,Yoo2010hto,Pomeau1994doa,Rakic2024epa,Poli2024eoa}
\begin{equation}
   \mu=\rho_n v_s^2\equiv(1-f_s)\rho v_s^2\,.
   \label{eqn:mu}
\end{equation}
In this expression, $\rho=mn_\text{2D}$ is the sum of the normal and superfluid densities, $\rho = \rho_n + \rho_s$, as defined by the superfluid fraction $f_s = \rho_s/\rho$ \cite{Leggett1970cas}. We calculate the superfluid fraction tensor,
\begin{equation}
   f_{ij} = \delta_{ij} - \frac{1}{\rho A}\frac{\partial P_i}{\partial v_j}\,, 
   \label{eqn:fs}
\end{equation}
with $i=x,y$, through the linear momentum response $\textbf{P}=(P_x,P_y)$ of a superfluid within a unit cell of area $A$ pushed at velocity $\textbf{v}=(v_x,v_y)$, using the auxiliary function approach introduced in Ref.~\cite{Blakie2024sft}. For the triangular and honeycomb states, the off-diagonal elements are consistent with zero, and we take $f_{ii}=f_s$ \footnote{As the superfluid fraction always exceeds $f_s=0.1$ we are always in the supersolid phase, though we have verified our results hold in the isolated droplet regime $f_s<0.1$.}. The physical interpretation of the superfluid fraction tells us the proportion of the state that reacts like a superfluid to an external perturbation. Experimentally, the superfluid response has been probed in 1D \cite{Tanzi2021eos,Biagioni2024mot} and 2D \cite{Norcia2022cao} crystals, and recent theoretical work suggests that $f_s$ can be extracted from condensate density images alone \cite{Blakie2024sft}.

\begin{figure}
    \centering
    \includegraphics[width=1\linewidth]{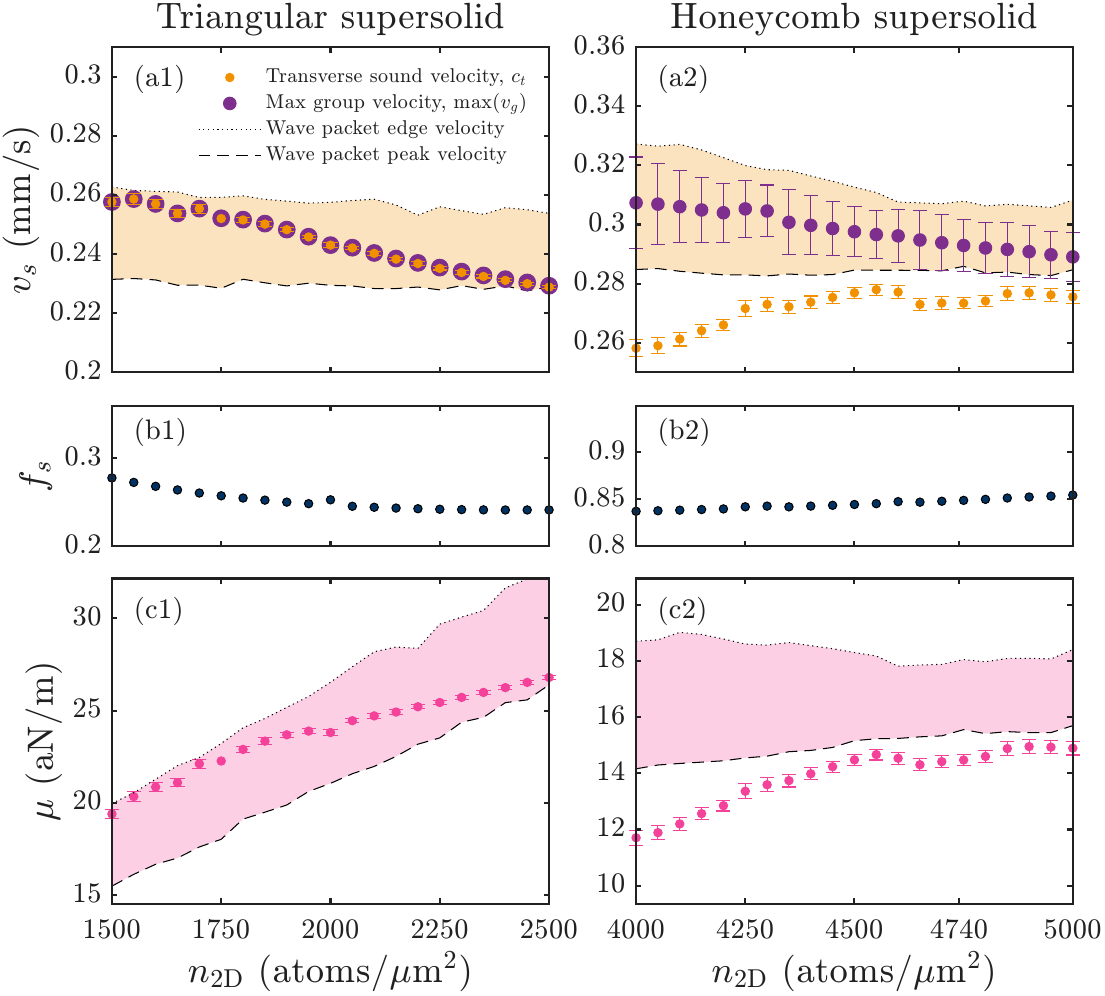}
    \caption{Shear wave velocities and the shear modulus for triangular $a_s/a_\text{dd} = 0.75$ and honeycomb $a_s/a_\text{dd} = 0.775$ supersolids. (a) Orange points indicate the estimated transverse speed of sound, $c_t$, and the purple points indicate the maximum value of the group velocity, $v_g$, with 1$\sigma$ confidence error bars from a quartic fit to the BdG calculations of the transverse branch. Shaded region is bounded by the slowest peak COM velocity (dashed curve) and edge COM velocity (dotted curve). (b) Superfluid fraction. (c) Shear modulus, $\mu$, calculated with $c_t$ and the COM velocities from} (a).
    \label{fig:shear}
\end{figure}

Fig.~\ref{fig:shear} shows the required ingredients in order to estimate the shear modulus from our measured time-dependent velocities $v_s$. We generalize our results to a wide-range of 2D densities, whilst maintaining the ground state crystal structure. The shaded region in (a) demarcates the range of velocities that could be observed in an experiment. The lower bound is given by the initial peak velocity (which increases over time), and the upper bound is the edge velocity. Combined with the superfluid fraction calculation of (b), we show that the shear moduli of both states are on the order of attoNewtons per meter (aN/m) \footnote{As the crystal is 2D, the units are in Newtons per length.}, and are consistent with recent related work~\cite{Poli2024eoa,Blakie2025}. Thus, the small--but non-zero--values of the shear moduli definitively show that the dipolar supersolid phases are solid \footnote{We have verified for both states that these observations are only weakly dependent on the scattering length.}.

Consider the experimental implementation of this protocol for the triangular supersolid. The edge velocity matches closely with our extracted $c_t$ [Fig.~\ref{fig:shear}(a1)], however due to fluctuations such as temperature effects, tracking the edge of the shear wave may be below the fidelity of experiments. Furthermore, in finite size systems the propagation velocity can be greatly reduced at the boundary of the system, as we show in the Supplemental Material~\cite{supp}. Thus, experimental estimates may drastically underestimate the shear modulus. At much larger densities than shown here, the transverse branch becomes soft, and the shear wave velocity tends to zero. However, this is in the parameter regime where the stripe phase is the ground state solution \cite{Ripley2023tds}.

The honeycomb state in Fig.~\ref{fig:shear}(a2) shows a propagation velocity exceeding $c_t$ across the parameter space, however this effect is diminished for increasing densities as indicated by the decreasing width of the shaded region. This transition from anomalous to normal dispersion is reminiscent of the behavior seen in superfluid helium-4 as a function of pressure \cite{sridhar1987apd}. The anomalous dispersion is maximized at low densities, close to the honeycomb-to-stripe transition, and may be related to the hybridization of modes across this structural phase transition. Refs.~\cite{Andreev1969qto,Yoo2010hto,Pomeau1994doa,Rakic2024epa,Poli2024eoa} define Eq.~\eqref{eqn:mu} using the transverse speed of sound, replacing $v_s\rightarrow c_t$. The prediction of anomalously dispersive transverse waves calls this definition into question. Instead, the stiffness is a $k$-dependent quantity, and the shear modulus of the honeycomb supersolid can exceed the one calculated from the speed of sound.

{\it Conclusions}---We have demonstrated that dipolar supersolids with two-dimensional crystalline structure are true solids displaying a finite shear modulus. However, the triangular and honeycomb states exhibit shear wave phenomena of a diverging nature. The triangular lattice structure demonstrates normal dispersion, and this modulated superfluid state behaves as an archetypal solid. In contrast, supersolids with honeycomb lattice structure show anomalous dispersion, which supports shear wave propagation exceeding the transverse speed of sound and opens intriguing avenues for future theoretical and experimental exploration. We expect a rich phenomenology to arise in this phase from the novel scattering and spontaneous decay channels of anomalously dispersive transverse phonons.

Our results are within current experimental capabilities. Triangular lattice dipolar supersolids are now routinely available \cite{Norcia2021tds,Bland2022tds,Casotti2024oov}, and a new generation of experiments with large dipolar atom numbers \cite{Krstajic2023cot,Anich2024cco} (or molecules \cite{Bigagli2024oob}) has brought the honeycomb state within reach \cite{Hertkorn2021pfi,Zhang2021pos,Schmidt2022sbd,Gallemi2022spo}. Once a 2D state has been achieved, a transverse wave packet can be excited through a local perturbation of the crystal structure. To provide the initial shift to the droplet positions, one could employ an optical lattice \cite{Halperin2023fia} or a digital micromirror device \cite{Gauthier2016dio}. The transverse branch can also be directly probed by imprinting transverse phonons, and the resulting structure factor measured using Bragg spectroscopy \cite{guo2021aol}.

Our work opens up new avenues for non-invasive measurements of dipolar supersolids. In future work, we will explore how defects in the supersolid affect the propagation of shear waves, bringing the ideas of shear wave elastography to the quantum realm. A supersolid can host both solid and superfluid defects, as holes in the crystal structure and quantized vortices, respectively. These superfluid vortices lie in the low-density regions, and are not possible to image directly \cite{Casotti2024oov,poli2024synchronization}. How these defects distort and interact with shear waves \cite{Son2005ela} may provide an indirect measurement of their presence in future dipolar supersolid experiments. Our findings may also provide insights into the propagation of shear waves in the supersolid phase of a neutron star following a starquake or glitch \cite{Poli2023gir}.  

{\it Acknowledgements}---We acknowledge invaluable conversations with F. Ferlaino, R. Bisset, E. Poli, P. B. Blakie, N. Liebster, M. Olimpo, and the Innsbruck dipolar team. We are particularly grateful to P. B. Blakie for sending us data from Refs.\,\cite{Ripley2023tds} and \cite{Poli2024eoa} to benchmark our results. We acknowledge financial support by the ESQ Discovery programme (Erwin Schrödinger Center for Quantum Science \& Technology), hosted by the Austrian Academy of Sciences (ÖAW), and the European Research Council through the Advanced Grant DyMETEr (10.3030/101054500). T.B. acknowledges funding from FWF Grant No.~I4426.

\end{document}


\title{Supplementary Material: Anomalous dispersion of shear waves in dipolar supersolids}

\author{P. Senarath Yapa}
\email{pramodh.sy@gmail.com}
\affiliation{
Universität Innsbruck, Fakultät für Mathematik, Informatik und Physik,
Institut für Experimentalphysik, 6020 Innsbruck, Austria
}

\author{T. Bland}
\email{thomas.bland1991@gmail.com}
\affiliation{
Universität Innsbruck, Fakultät für Mathematik, Informatik und Physik,
Institut für Experimentalphysik, 6020 Innsbruck, Austria
}

\maketitle

\section{Extended Gross-Pitaevskii Theory}\label{sec:theory}

To model our infinite planar system, the 3D wavefunction can be separated into $\Psi(\textbf{r},t)\equiv\Psi(\boldsymbol{\rho},z,t)=\psi(\boldsymbol{\rho},t)\chi_\sigma(z)$, where $\boldsymbol{\rho}=(x,y)$ and $\chi_\sigma(z) = \pi^{-1/4}\sigma^{-1/2}e^{-z^2/2\sigma^2}$ with variational width $\sigma$. Following Ref.~\cite{Ripley2023tds}, we substitute this ansatz into the 3D eGPE energy functional and perform the integration over $z$ to find the new total energy functional $E=E_\perp+E_z$, where
\begin{align}
    E_\perp = \int\text{d}^2\boldsymbol{\rho}\,\psi^*\bigg[-\frac{\hbar^2\nabla^2}{2m} +\frac12\Phi  +\frac25g_\text{QF}|\psi|^3\bigg]\psi\,,
    \label{eqn:energy_perp}
\end{align}
and $E_z=\hbar\omega_z(l_z^2/4\sigma^2 +\sigma^2/4l_z^2)$, where $l_z = \sqrt{\hbar/m\omega_z}$ is the harmonic oscillator length. The two-dimensional interaction potential can be evaluated in Fourier space as 
\begin{align}
    \Phi(\boldsymbol{\rho},t) = \mathcal{F}^{-1}\left[\tilde{U}(\boldsymbol{k})\mathcal{F}\left[|\psi(\boldsymbol{\rho},t)|^2\right]\right]\,,
\end{align}
with $\boldsymbol{k}=(k_x,k_y)$, (inverse) Fourier transform $\mathcal{F}$ ($\mathcal{F}^{-1}$), and 
\begin{align}
    \tilde{U}(\boldsymbol{k}) = \frac{\sqrt{2\pi}\hbar^2}{\sigma m}\left(a_s+a_\text{dd}\left[2-3G_0\left(\sigma \boldsymbol{k}/\sqrt{2}\right)\right]\right)\,,
\end{align}
and $G_0(q) = \sqrt{\pi}qe^{q^2}\text{erfc}(q)$, is the dimensionally reduced Eq.\,\eqref{eqn:U}.
Finally, we have also introduced the following scaled quantum fluctuation term \cite{Schuetzhold2006mfe,Lima2011qfi}
\begin{align}
    g_\text{QF} = \frac{128\hbar^2}{3m}\sqrt{\frac{2a_s^{5}}{5\pi^{1/2}\sigma^3}}\, \,\left[1+\frac{3}{2}\left(\frac{a_{\mathrm{dd}}}{a_s}\right)^2\right]\,.
\end{align}
One can calculate $\delta E_\perp/\delta\psi^*$ to obtain the quasi-2D variational eGPE
\begin{align}
    i\hbar\frac{\partial\psi}{\partial t} =  \bigg[-\frac{\hbar^2\nabla^2}{2m} + \Phi_\sigma+g_\text{QF}|\psi|^3 \bigg]\psi\,.
    \label{eqn:2DGPE}
\end{align}

\begin{figure}
    \centering
    \includegraphics[width=1\columnwidth]{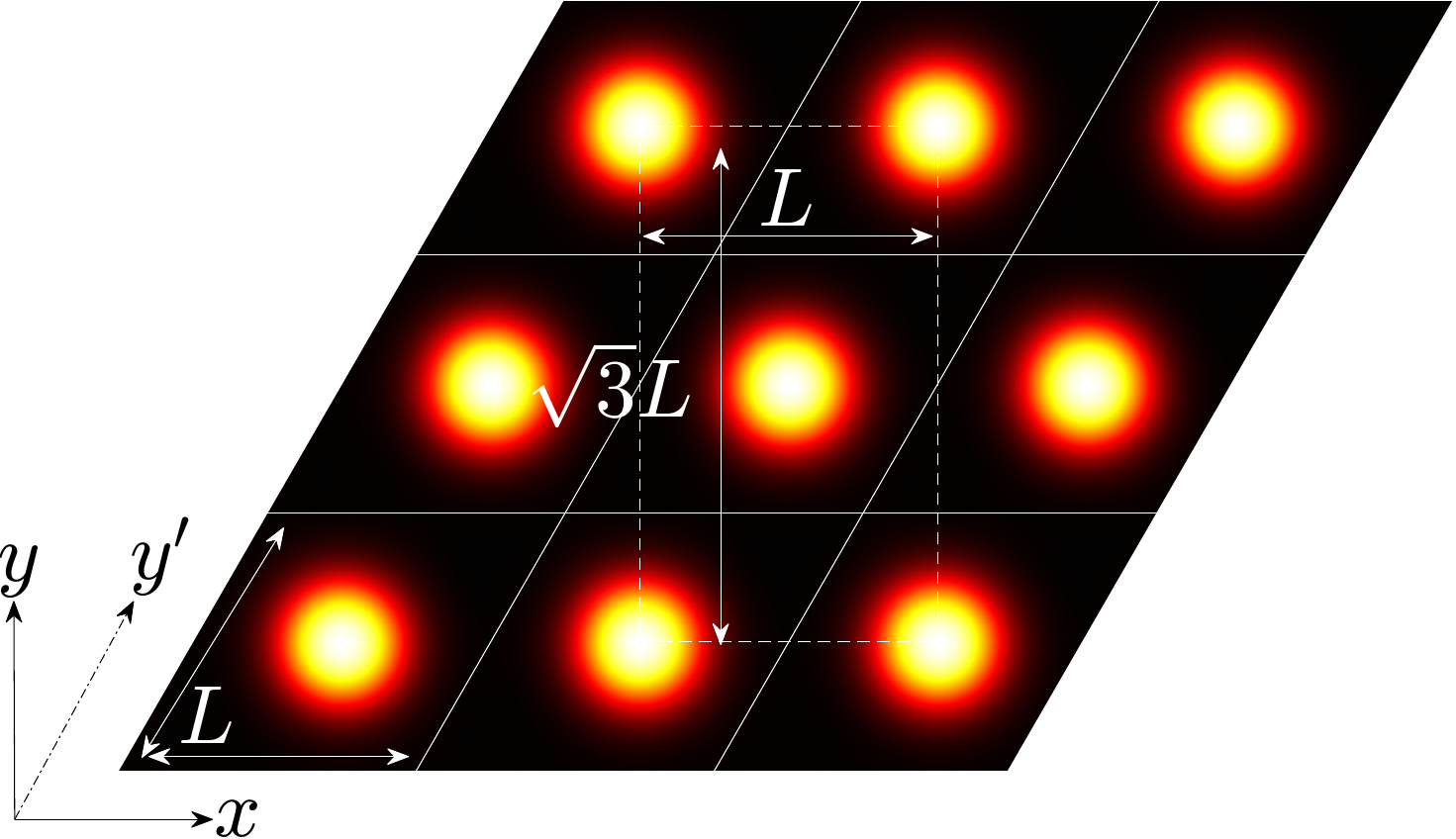}
    \caption{Initial state construction, see Sec.~\ref{sec:theory}.}
    \label{fig:shear_eg}
\end{figure}

Fig.~\ref{fig:shear_eg} shows an exemplar triangular density distribution obtained with our model, and below we list the steps we take to construct the full state. This solution is found by first solving Eq.\,\eqref{eqn:2DGPE} for a given scattering length and 2D density in imaginary time ($t\to-i\tau$) using a split-step Fourier method. We take a parallelogram unit cell with $[N_{x},\,N_{y^\prime}] = [72,\,72]$ grid points. We minimize the total energy $E$ with respect to the variational parameter $\sigma$, the wavefunction, and the length $L$ of the parallelogram unit vectors. This procedure has been covered in detail in Refs.\,\cite{Ripley2023tds,Ripley2023soa}. We then calculate the superfluid fraction $f_s$, Eq.~(6) of the main text, from this ground state as described in Ref.\,\cite{Blakie2024sft}. We tile this solution in a $3\times3$ grid, and extract a rectangular cell (with Cartesian coordinates) of size $L\times\sqrt{3}L$ (dashed line). Finally, the rectangular cell is tiled to cover a large area of $2L\times 120\sqrt{3}L$ with $[N_{x},\,N_{y}] = [80,\,7200]$ grid points, forming our initial state to perturb and investigate shear wave propagation.

\begin{figure}
    \centering
    \includegraphics[width=1\columnwidth]{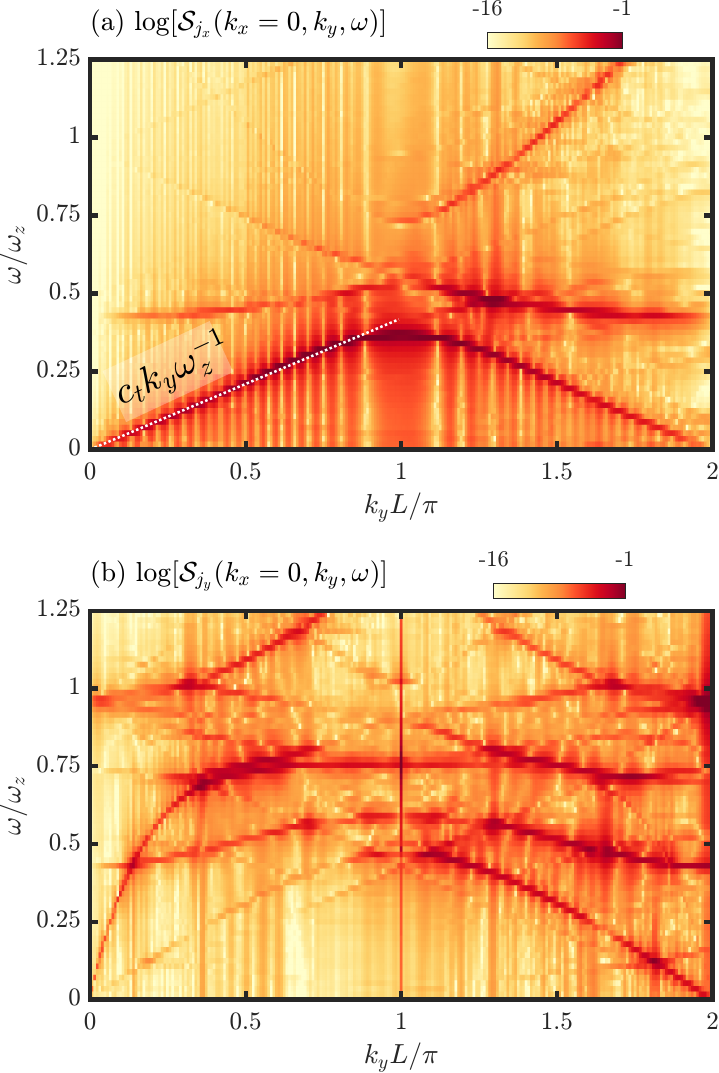}
    \caption{Current density DSFs for the honeycomb state presented in Fig.~2 of the main text, with (a) $\mathcal{S}_{j_x}$ and (b) $\mathcal{S}_{j_y}$. The dotted line in (a) shows the low-$k$ linear dispersion, with gradient equal to the speed of transverse sound.}
    \label{fig:honeycomb_DSF}
\end{figure}

\section{Extraction of Dynamic Structure Factor}

In order to create panel (c) in Figs. 1 and 2 of the main text, we take the difference of the two current density DSFs $\mathcal{S}_{j_x}$ and $\mathcal{S}_{j_y}$. Here, we wish to highlight the data post-processing required to reduce the noise on this figure, and to find the superfluid, longitudinal crystal, and transverse crystal phonon branches.

In Fig.~\ref{fig:honeycomb_DSF} we show the raw data of the two current density DSFs of Fig.~2(c) on a logarithmic colorscale. Through the numerical noise, the three Goldstone branches are clearly visible. On the transverse branch, we overlay the linear approximation $\omega = c_tk_y$, showing that the branch visibly exceeds this initially linear trajectory. Other features of interest can be discerned from this figure. The first gapped branch (around $\omega=0.5\omega_z$) is brightest in the second BZ, and corresponds to a sub-lattice wavelength shearing of the holes of the honeycomb. A similar feature is shown in (b), where the superfluid phonons of sub-lattice wavelength  also couple strongest to this excitation scheme. When plotting Fig.~2 we isolate the peaks of this data, throwing away signals of amplitude $<0.01$ to highlight the relevant branches.

\begin{figure}
\vspace{0.3cm}
    \centering
    \includegraphics[width=1\columnwidth]{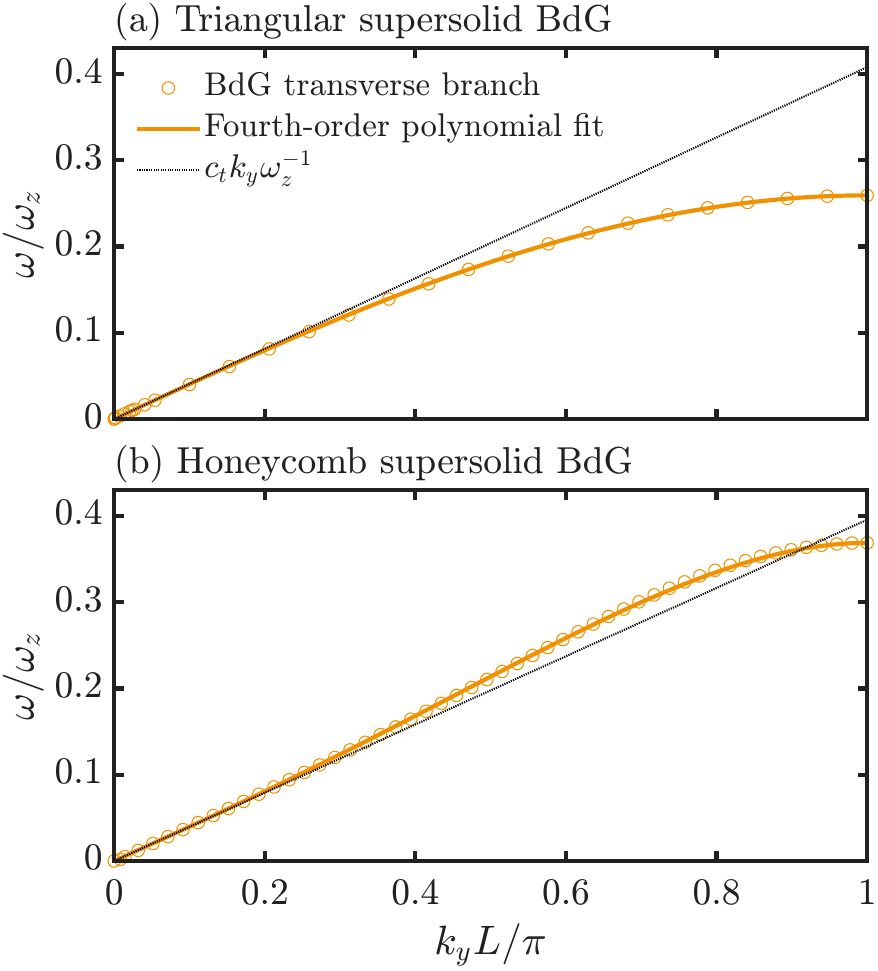}
    \caption{The transverse branch obtained from the BdG calculations for the (a) triangular supersolid in Fig.~1(c) and (b) honeycomb supersolid in Fig.~2(c). The dotted lines show the low-$k$ linear dispersion, with gradient equal to the speed of transverse sound.}
    \label{fig:transvbranch}
\end{figure}

\section{Bogoliubov-de Gennes Theory}

To validate the dispersion obtained from the current density DSFs, we expand the Bogoliubov modes in Eq.~(4) of the main text into Bloch waves
\begin{equation}
u_\nu(\boldsymbol{\rho})=u_{\nu, \boldsymbol{k}}(\boldsymbol{\rho}) e^{i \boldsymbol{k} \cdot \boldsymbol{\rho}}, \quad v_\nu(\boldsymbol{\rho})=v_{\nu, \boldsymbol{k}}(\boldsymbol{\rho}) e^{i \boldsymbol{k} \cdot \boldsymbol{\rho}}.
\end{equation}
By substituting the perturbation into the quasi-2D eGPE and linearizing, we obtain the BdG matrix
\begin{equation}
\hbar\omega_\nu(\boldsymbol{k})
\begin{bmatrix}
u_{\nu,\boldsymbol{k}} \\
v_{\nu,\boldsymbol{k}}
\end{bmatrix}
=
\begin{bmatrix}
\mathcal{L}+\mathcal{M} & -\mathcal{M} \\
\mathcal{M} & -\left(\mathcal{L}+\mathcal{M}\right)
\end{bmatrix}
\begin{bmatrix}
u_{\nu,\boldsymbol{k}} \\
v_{\nu,\boldsymbol{k}}
\end{bmatrix},
\end{equation}
where the operators are
\begin{align}
\mathcal{L} &= -\frac{\hbar^2\nabla^2}{2m} -\mu + \Phi_\sigma+g_\text{QF}|\psi|^3,\\
\mathcal{M}f &= \psi_0\mathcal{F}^{-1}\left[\tilde{U}(\boldsymbol{k}) \mathcal{F}\left[f \psi_0^*\right]\right]+\frac{3}{2} g_{\mathrm{QF}}|\psi_0|^3 f.
\end{align}
We fit the transverse branch obtained from this procedure with a quartic polynomial in order to extract the three characteristic velocities shown in panel (d) in Figs.~2 and 3. In Fig.~\ref{fig:transvbranch} we show the transverse branches and the quartic fit for the triangular and honeycomb supersolids at $n_\text{2D} = 2000\,\mu$m$^{-2}$ and $n_\text{2D} = 4300\,\mu$m$^{-2}$ respectively. Similar to Fig.~\ref{fig:honeycomb_DSF}, we overlay the linear approximation $\omega = c_tk_y$, showing that the transverse branch of the honeycomb supersolid exceeds this linear trajectory in Fig.~\ref{fig:transvbranch}(b). 

\section{Shear waves in a finite size system}

It is pertinent to ask how our results scale in a finite sized system. As the elastic tension for droplets along the edge of the system are only provided by a single neighboring row, can shear waves still propagate in this system? How does the inhomogeneity affect our results? We tackle these questions by considering a system with an additional harmonic trap along one axis, i.e. inserting $V = m\omega_y^2y^2/2$ into our Hamiltonian in Eq.~\eqref{eqn:2DGPE}, such that our system emulates a segment of a supersolid in a ring trap geometry \cite{Tengstrand2021pci,Tengstrand2023tds,Hertkorn2024dsa}. 

\begin{figure}[b]
    \vspace{0.3cm}
    \centering
    \includegraphics[width=1\columnwidth]{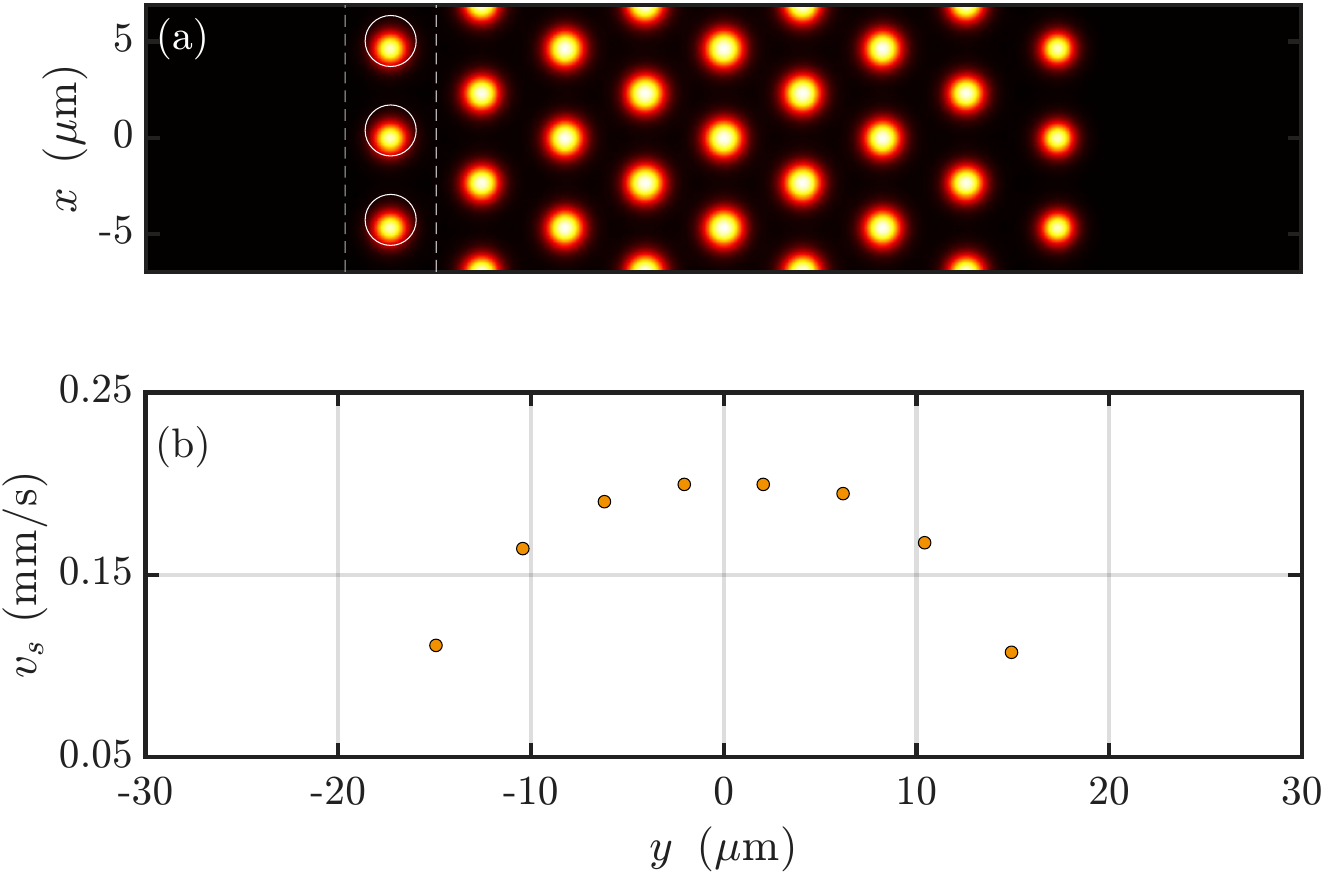}
    \caption{Propagation of a shear wave in a triangular lattice dipolar supersolid harmonically trapped in both $y$ and $z$ with $(\omega_y,\,\omega_z) = 2\pi\times(15,\,72.4) \mathrm{~Hz}$. Ground state results for $a_s/a_\text{dd} = 0.750$ and $N = 5.5\times10^5$ atoms. (a) The initial density is locally perturbed along $\hat{x}$ (between the dashed lines), resulting in the propagation of a shear wave packet along $\hat{y}$. The white circles show the initial droplet positions. (b) The velocity of the shear wave measured locally using the time delay at which pairs of columns reach their maximum displacement.}
    \label{fig:shear_finite}
\end{figure}

Fig.~\ref{fig:shear_finite} shows our results. The initial state, in (a), is excited via a vertical shift applied to the outermost column of droplets. We find that, as in the infinite supersolid, a shear wave propagates across the system. Unlike the infinite case, there is now a spatially varying velocity, peaking where the density is largest. We define the local velocity between the columns by the time delay at which pairs of columns reach their maximum displacement. This change in velocity is due to the variation in the inter-droplet spacing between columns. Our results verify that shear waves can propagate in an experimentally accessible geometry. In future work, we will investigate this phenomena in two-dimensional box-traps, where the trap itself will play less of a role \cite{Roccuzzo2022sea}, or in azimuthally symmetric harmonic traps, where the radial shear waves are Tkachenko modes \cite{Tkachenko1966ovl,Tkachenko1966sov,Tkachenko1969eov,Coddington2003oot}.

\sloppy